\documentstyle[preprint,aps]{revtex}
\begin{document}
\draft
\title{\hspace{25ex} {\small\sc Applied Physics Report 94-13} \\
\hspace{25ex} {\small\sc  cond-mat/9404077} \\
Soft Disorder Effects in the Conductance Quantization in Quantum Point
Contacts: Indirect Backscattering Statistics}
\author{ Alexandre M. Zagoskin\thanks{Email:alexz@fy.chalmers.se}, Sergey N.
Rashkeev\thanks{Email:sergey@fy.chalmers.se}, Robert I.
Shekhter\thanks{Email:shekhter@fy.chalmers.se}, and G\"{o}ran
Wendin\thanks{Email:wendin@fy.chalmers.se}\\
Department of Applied Physics, Chalmers University of Technology and University
of G\"{o}teborg, S-41296 G\"{o}teborg, Sweden }
\maketitle

\begin{abstract}
The breakdown of conductance quantization in a quantum point contact
in the presence of random long-range impurity potential is discussed.
It is shown that in
the linear response regime  a decisive role is played by the  indirect
backscattering mechanism via quasilocalized states at the Fermi level; this can
provide much higher backscattering rate than any direct backscattering process.
For the realistic contact lengths ($\leq 2000$nm) the  scattering
processes prove to be independent, in spite of coherence of the electron wave.
The  distribution function  of conductance fluctuations is obtained by direct
numerical calculations as well as estimated within an analytical model for
the first time. It is
shown to be a generalized Poisson distribution. Estimates are made for  quantum
point contact performance at different choice of parameters. In particular,
it is the better the larger the intermode  distance is  compared to the
amplitude of the random impurity potential.
\end{abstract}

\section{Introduction}\label{s.0}

The effect of conductance $2e^{2}/h$-quantization in quantum point contacts
(QPC, electrostatically defined junction in a two-dimensional electron gas in a
high mobility GaAs-AlGaAs heterostructure), first observed in 1988 \cite{WW},
is still challenging both theoreticians ant experimentalists \cite{OBZOR}. An
intriguing feature of the effect is  that it is not destroyed by elastic
scattering,  be it impurity scattering or scattering by a contact boundary.

The insensitivity of  quantization  to the latter process was explained within
the framework of the adiabatic approximation \cite{AA}. Breakdown of
quantization was shown to be insignificant  as long as the contact shape (i.e.,
the confining potential induced by the gate electrode) is smooth on the scale
of Fermi wavelength, $\lambda_{F}$. The last condition is likely to be true,
since  $\lambda_{F} \simeq 40$nm in a clean contact is at least an order of
magnitude less than the characteristic scale of the gate  potential variation
\cite{Davies}.

The adiabatic approximation yields that different transverse modes
pass the QPC  independently ("no-mode-mixing" regime). This provides a
sufficient,
but not necessary, condition for the conductance quantization in a QPC: as was
pointed in \cite{Davies,Buttiker1}, the necessary condition is the absence of
backscattering. The direct numerical calculations \cite{Chao} show that
conductance is quantized even if the intermode mixing is significant.  The sum
rule suggested there to explain this result follows from the unitarity of the
scattering matrix of the system \cite{ZS}.

The impurity  potential is more likely to produce backscattering, and thus to
break the conductance quantization, since it is less smooth than the
gate-induced potential. Many theoretical papers dealing with   impurity
scattering in QPC \cite{DasSarma,Maslov}  used the Anderson model with on-site
disorder. This approach gives a qualitative understanding of the process, but
its results cannot be  directly applied to  real GaAs planar structures, where
the Coulomb impurity potential certainly  is not   sharp on the scale of
$\lambda_{F}$ \cite{Davies,GJ}.  In fact, the numerical calculations made for
this realistic case \cite{Davies} show that its scale of variation  is
intermediate between $\lambda_{F}$ and the size of the  QPC itself.

The effect of such a slowly varying random potential ("soft disorder") on the
performance of a QPC was investigated in \cite{Davies,GJ}. Its main feature is
the negligible rate of scattering processes with large momentum transfer.
Glazman and Jonson \cite{GJ} built a theory of the electrical conductivity
through the QPC with soft disorder and showed that the {\em direct
backscattering} occurs mainly in the largest-number transverse mode (i.e., one
with the minimal longitudinal momentum).  On another hand, Laughton {\em et
al.} \cite{Davies} demonstrated that the long range impurity potentials may
produce quasilocalized states  inside the channel \cite{FN0}. Once appeared,
such a state provides an effective {\em indirect} backscattering mechanism.
Indeed, the quasilocalized state contains both "forward" and "backward"  waves,
so that the transition  between the electronic states propagating in  opposite
directions via the quasilocalized state  does not demand a significant momentum
transfer. Thus, in an appropriate impurity configuration, indirect
backscattering  (formally due to the  second order process) can have much
larger cross section than the (first order) direct backscattering. The
probability to find such a configuration of impurities, of course, must be
investigated as well. Recently Gurvitz and Levinson \cite{Lev} built a theory
of resonant reflection and transmission in a QPC with a single attracting
impurity at the contact's bottleneck. They have investigated the case when  the
first mode is open and a second  one is being opened, thus creating a single
resonant level at the top of the potential well. They  obtained a Breit-Wigner
type expression for the correction to the conductance at the transition region
from the first to second conductance step (as function of the Fermi energy);
its sign   depends on whether there takes place tunneling into the resonant
state (conductance  enhancement) or scattering into it (suppression).

In this paper we study numerically and analytically the indirect backscattering
effects on the electrical conductivity of a QPC, in the presence of the
screened Coulomb potential from randomly distributed charged impurities.

In Sec.\ref{s.1} we  find the corresponding correction to the current,
supposing that there are quasilocalized states in the QPC, and that the
impurity potential  is soft, so that the tunneling processes to and from these
states can be neglected, and their energetic spectrum is dense. We show that
the correction, given by a Breit-Wigner type formula, suppresses the current.
If the quasilocalized states exist close to the Fermi surface,  this is  really
the leading contribution from the impurity scattering in a QPC. The result is
valid for the bulk of the conductance step vs. gate voltage.

In Sec.\ref{s.2} we investigate the stochastic properties of arising
conductance fluctuations, as a random variable determined by the specific
impurity arrangement.  The distribution function of  the conductance
fluctuations is shown to be of Poisson type and is explicitly determined by the
tunable parameters of the QPC: its length and number of open modes (i.e.,
conductance).
This allows us to make more detailed predictions about the QPC performance than
mere knowledge of the fluctuation dispersion.

In Sec.\ref{s.3} we estimate the theory parameters based on the quasiclassical
description of the random impurity potential.

In Sec.\ref{s.4}  we  present the numerical calculations  that provide the
basis for the present work. The impurity potential was calculated  in a
self-consistent way using the Thomas- Fermi approximation.  The quantum contact
was modelled by imposing upon it a parabolic confining potential. The
conductance was calculated in the linear responce limit by a standard transfer
matrix method for different contact lengths and  for different realizations of
the impurity potential. ( Different  realizations of the random  impurity
potential in the QPC was performed by changing the position of the confining
potential with respect to once obtained random potential profile.)

An empirical distribution function of conductance fluctuations was thus
obtained, and we  compare the results of the numerical calculations and
theoretical predictions in Sec.\ref{s.A}. They are in very good agreement. This
allows us to conclude that the leading mechanism of conductance quantization
breakdown in QPCs is the indirect backscattering, and it allows us to find
estimates of the QPC performance.

\section{Indirect backscattering in a quantum point contact}\label{s.1}

In accordance with what is now the standard approach \cite{AA,GJ}, we start
from the adiabatic model of a QPC, where the transverse modes are well defined;
the deviations will be regarded as a perturbation (see Fig.\ref{f.1}). The
Hamiltonian of the electron contains the following terms:
\begin{eqnarray}
H = -\frac{\hbar^{2}}{2m^{*}} \left( \frac{\partial^{2}}{\partial x^{2}} +
\frac{\partial^{2}}{\partial y^{2}} \right) + U_{g}(x,y) + V(x,y).
\end{eqnarray}
Here $U_{g}(x,y)$ is the confinement potential, shaping the junction in a
two-dimensional electron gas (2DEG). It proves to be convenient to include
the smooth component of the impurity potential as well. The rest of impurity
potential, $V(x,y)$, is regarded as a perturbation, e.g., leading to intermode
mixing. It is also a soft potential, which does not lead to  large momentum
transfer.

The potential $U_{g}(x,y)$ is supposed to be a slow function of the
coordinates. Then the wave function of an electron with energy $E$ inside the
QPC  can be expanded as follows \cite{AA}:
\begin{eqnarray}
\Psi(x,y;E) = \sum_{m,\alpha} C_{m,\alpha}\psi_{m,\alpha}(x,y;E).
\end{eqnarray}
The adiabatic eigenfunctions are defined by
\begin{eqnarray} \psi_{m,\alpha}(x,y;E)   = \phi_{m}(x,y) \chi_{m,\alpha}(x;E),
\end{eqnarray}
where the transverse eigenfunction corresponds to the $m$-th eigenvalue of the
transverse Hamiltonian:
\begin{eqnarray}
H_{\perp}\phi_{m}(x,y) \equiv \left\{ -\frac{\hbar^{2}}{2m^{*}}
\frac{\partial^{2}}{\partial y^{2}}  + U_{g}(x,y)\right\}\phi_{m}(x,y) =
 E_{m\perp}(x)\phi_{m}(x,y),
\end{eqnarray}
and the longitudinal eigenfunction satisfies the equation
\begin{eqnarray}
\left\{ -\frac{\hbar^{2}}{2m^{*}} \frac{\partial^{2}}{\partial x^{2}}  +
E_{m\perp}(x) \right\}\chi_{m,\alpha}(x;E) =
E \chi_{m,\alpha}(x;E).
\end{eqnarray}

We can distinguish different groups of electronic states in the QPC, according
to the behaviour of $\chi_{m,\alpha}(x;E)$ in the presence of the effective
one-dimensional potential $E_{m\perp}(x)$ (see Fig.\ref{f.2}). They are denoted
by the index $\alpha = -2,-1,.. 2$.

The states of special interest for us are the quasilocalized ones ($\alpha=0$).
They appear if the effective potential $E_{m\perp}(x)$ is a nonmonotonous
function. Note that for each quasilocalized state in $m$-th mode, generally,
there exist propagating states with the same energy in some mode of less number
  (Fig.\ref{f.2}). This is the reason why we  these states are {\em
quasi}localized (due to smoothness of the potentials in the QPC we can safely
neglect another reason for this, namely, the tunneling decay).

The coexistence  of propagating and quasilocalized states (P-states and
Q-states) at the same energy  is characteristic for QPCs and gives rise to
indirect backscattering  in the QPC.

The  correction to the current  due to backscattering can be written in a
standard way \cite{AA,Imry}:
\begin{eqnarray}
\Delta I = -\frac{2|e|}{h} \int dE \left(n_{F}(E-\mu_{1}) -
n_{F}(E-\mu_{2})\right) J(E); \label{current}\\
\nonumber\\
J(E) = \sum_{m=1}^{N}\sum_{n=1}^{N} \left|r_{nm}(E)\right|^{2}.
\label{current2}
\end{eqnarray}
In the linear response limit this yields the correction to the conductance,
\begin{eqnarray}
\Delta G = -\frac{2e^{2}}{h} J(E_{F}). \label{conductance}
\end{eqnarray}
The summation  in (\ref{current2}) is taken only over $N$  propagating (open)
modes  in the quantum point contact, and $r_{nm}(E)$ is the probability
amplitude of the electron incident from the right in $m$th P-mode ($\alpha=1$)
to be  scattered back, to the $n$th P-mode with ($\alpha=-1$):
\begin{eqnarray}
r_{nm}(E)\delta(E-E') = \left(\psi_{n,-1}(E'), \Psi^{out}_{m,1}(E) \right).
\label{R}
\end{eqnarray}
Here $\Psi^{out}_{m,1}(x,y;E)$ is the  scattered wave, corresponding to the
unit wave with energy $E$, incident from the right in the $m$th mode,
$\psi_{m,1}(x,y;E)$.  This function can be found from the Lippmann-Schwinger
equation \cite{SCAT} (${\bf r} = (x,y)$):
\begin{eqnarray}
\Psi^{out}_{m,1}({\bf r};E) = \psi_{m,1}({\bf r};E) +  \int d{\bf r'}
G^{R}({\bf r,r'};E)V({\bf r'})
\psi_{m,1}({\bf r'};E). \label{LS}
\end{eqnarray}
In this equation we have introduced the exact retarded Green's function
(retarded Green's function) in the presence of the perturbation $V(x,y)$, for
which we can   write :
\begin{eqnarray}
G^{R}({\bf r,r'};E) = G^{R}_{0}({\bf r,r'};E) +
 \int d{\bf r''} G^{R}_{0}({\bf r,r''};E)V({\bf r''})G^{R}_{0}({\bf r,r'};E) +
... \label{DE}
\end{eqnarray}
The unperturbed retarded Green's function is given by
\begin{eqnarray}
G^{R}_{0}({\bf r,r'};E) = \sum_{m,\alpha,\epsilon} \frac{\psi_{m,\alpha}({\bf
r};\epsilon)\psi_{m,\alpha}^{*}({\bf r'};\epsilon)}{E - \epsilon + i0} \equiv
\sum_{\alpha} G^{R; \alpha}_{0}({\bf r,r'};E). \label{retarded Green's
function}
\end{eqnarray}
The summation is taken over all the quantum numbers   of electronic unperturbed
eigenfunctions.

Inserting (\ref{LS},\ref{DE}) into (\ref{R}) we find  two first nonvanishing
terms:
\begin{eqnarray}
r_{nm}(E)\delta(E-E') = \left(\psi_{n,-1}(E'),
G^{R}_{0}(E)V\psi_{m,1}(E)\right) +
 \left(\psi_{n,-1}(E'), G^{R}_{0}(E)VG^{R}_{0}(E)V\psi_{m,1}(E)\right).
\label{ll}
\end{eqnarray}
 We neglect the higher-order terms which is equivalent to an assumption
that the scattering processes by different quasilocalized levels are
independent. This assumption is valid only in the case $L \ll l_{tr},l_{loc}$,
where $l_{tr},l_{loc}$ are the transport and localization lengths,
respectively.
The following results show that this  condition holds in the case of realistic
quantum contacts.

According to our initial assumptions, the perturbation potential is soft. This
means that, generally, its matrix elements between the P-states propagating in
opposite directions (i.e., with $\alpha = +1$ and $-1$) are negligible, while
the matrix elements between the P-states and Q-states ($\alpha = \pm1$ and $0$)
are   nonzero. Since the retarded Green's function (\ref{retarded Green's
function}) contains the products of electronic eigenfunctions with the same
indices $\alpha$, Eq.(\ref{ll}) reduces to:
\begin{eqnarray}
r_{nm}(E)\delta(E-E') =  \left(\psi_{n,-1}(E'),
G^{R;-1}_{0}(E)VG^{R;0}_{0}(E)V\psi_{m,1}(E)\right). \label{main1}
\end{eqnarray}
The superscripts in the retarded Green's functions show which part of the
expansion (\ref{retarded Green's function}) we keep.

The formula (\ref{main1}) shows that the backscattering from the incident
(m,1)-state  to the (n,-1)-state occurs through the set of quasilocalized
states (described by the part of retarded Green's function, denoted by
$G^{R;0}_{0}(E)$).

After some standard transformations, we obtain that
\begin{eqnarray}
r_{nm}(E) =
2\pi\nu(E)\sum_{q,\epsilon} \frac{
\left(n;-1;E\right|V\left|q;0;\epsilon\right)
\left(q;0;\epsilon\right|V\left|m;1;E\right) }{ E - \epsilon + i0}.  \label{l2}
 \end{eqnarray}
Here we have introduced the one-dimensional electronic density of states (DOS)
at infinity, $\nu(E)$, to account for the propagating modes,  and denoted the
matrix elements of the perturbation potential by
\begin{eqnarray}
\left(n;-1;E\right|V\left|q;0;\epsilon\right) \equiv \int d{\bf r}
\psi_{n,-1}^{*}({\bf r};E) V({\bf r}) \psi_{q,0}({\bf r};\epsilon).
\end{eqnarray}
The summation is now taken only over the quasilocalized states;  as such, they
have discrete spectrum, so that $\epsilon$ is a discrete variable; $q$ is the
number of transverse mode where the localized state appears.

The equation (\ref{l2}) contains the unperturbed retarded Green's function of
the electron in localized states, which thus have an infinite lifetime. The
perturbation, mixing different states, makes the Q-states metastable, and
instead of $i0$-term there appears the spectral  function
$i\Gamma_{q}(\epsilon,\epsilon)/2$\cite{SCAT} (the accompanying shift of the
energy levels can be accounted for by changing the summation variable
$\epsilon$). The spectral function is given by
\begin{eqnarray}
\Gamma_{q}(\epsilon,\epsilon') = 2\pi\sum_{n}\sum_{\alpha=\pm1} \left|
\left(q;0;\epsilon\right|V\left|n;\alpha;\epsilon'\right) \right|^{2}
\nu(\epsilon); \label{Gamma}
\end{eqnarray}
(of course, a Q-state can decay only into P-states).

The correction to the current (\ref{current}) is expressed through $J(E) =
\sum_{m,n}|r_{nm}(E)|^{2}$. Note that the matrix elements corresponding to the
Q-states from different subbands $(q \neq q')$ or different impurity potential
wells in the same subband $(w \neq w'$, where $w$ is the label of the
potential well) enter the expression for $r_{nm}(E)$  with their phases, which
are   uncorrelated. This means that the main contribution to $J(E)$ will be
given by the diagonal terms in the corresponding sum, that is, as it is easily
seen,

\begin{eqnarray}
J(E) \simeq 4\pi^{2} \nu^{2}(E) \times \nonumber \\
\sum_{q,w}\sum_{\epsilon,\epsilon'} \frac{\sum_{m,n}
\left(n;-1;E\right|V\left|q;0;\epsilon\right)
\left(q;0;\epsilon\right|V\left|m;1;E\right) \cdot
\left(n;-1;E\right|V\left|q;0;\epsilon'\right)
\left(q;0;\epsilon'\right|V\left|m;1;E\right) }{ \left(E - \epsilon +
i\Gamma_{q}(\epsilon,\epsilon)/2 \right) \left(E - \epsilon' -
i\Gamma_{q}(\epsilon',\epsilon')/2 \right)}. \label{l4-0}
\end{eqnarray}
Here $\epsilon, \epsilon'$ are different quasilocalized states in the {\em
same} potential well of the {\em same} 1D subband.

This expression can be further simplified, if the distance between the energy
levels in a potential well are large comparatively to their width: $\Delta
\epsilon_{well} \gg \Gamma$. Then in the sum over $\epsilon, \epsilon'$ in
(\ref{l4-0}) the terms with $\epsilon \neq \epsilon'$ are negligibly small, so
that finally we obtain:

\begin{eqnarray}
J(E) \simeq 4\pi^{2} \nu^{2}(E) \sum_{q,\epsilon}\frac{\sum_{m,n}
\left|\left(n;-1;E\right|V\left|q;0;\epsilon\right)
\right|^{2}\left|\left(q;0;\epsilon\right|V\left|m;1;E\right)\right|^{2} }{ (E
- \epsilon)^{2} + (\Gamma_{q} (\epsilon,\epsilon)/2)^{2}} \nonumber\\
 =\sum_{q,\epsilon} \frac{(\Gamma_{q} (\epsilon,E)/2)^{2}}{ (E - \epsilon)^{2}
+
(\Gamma_{q}(\epsilon,\epsilon)/2)^{2}}. \label{l4}
\end{eqnarray}

The result has a typical Breit-Wigner form for a set of independent resonant
levels \cite{SCAT}, which is the case under the assumptions made.

Though the   energy spectrum of  the Q-states in a single potential well is
assumed to be rarified, in a long contact, where many potential wells with
random parameters appear, and due to the fact that each Q-state has a finite
width $\Gamma$,  the spectrum of these states for the whole system is dense
enough to  be described in countinuous approximation.  Then we can  introduce
the density of localized states in the $q$th mode, $\aleph_{q}(\epsilon)$, and
rewrite (\ref{l4}) as
\begin{eqnarray}
J(E) = \sum_{q} \int d\epsilon
\frac{\aleph_{q}(\epsilon)(\Gamma_{q}(\epsilon,E)/2)^{2}}{ (E - \epsilon)^{2} +
(\Gamma_{q}(\epsilon,\epsilon)/2)^{2}} \simeq
\frac{\pi}{2} \sum_{q} \aleph_{q}(E)\Gamma_{q}(E,E). \label{approx1}
\end{eqnarray}

In  the linear response limit this directly gives the  correction to the
conductance:
\begin{eqnarray}
\Delta G \simeq -\frac{2e^{2}}{h} \frac{\pi}{2} \sum_q
\aleph_{q}(E_{F})\Gamma_{q}(E_{F},E_{F}). \label{approx2}
\end{eqnarray}


We  discussed only virtual (including resonant) scattering to and from the
Q-states. In reality, when the finite driving voltage $U$ is applied, there
exist the processes of {\em real} elastic and inelastic (electron-electron or
electron-phonon) scattering between Q- and P-states with energies in the
interval $eU$ around the Fermi energy. The contribution of these states to the
scattering is proportional to $eU$. We can safely neglect their existence in
the if the  linear response limit $eU \rightarrow 0$, when the contribution
from the Q-states outside the $eU$-band  (in the interval of width $\sim
\Gamma(E_{F},E_F)$) is dominant. These kinetic processes, of course, must be
taken into account if we would like to discuss the nonlinear response of a QPC.


\section{Statistics of conductance fluctuations}\label{s.2}

Now let us discuss what the formula (\ref{approx2}) yields. The conductance
quantization breakdown can be characterized by the relative   difference
between the actual  conductance  vs. gate voltage dependence,  $G(V_{g})$, and
the ideal one, averaged over the $n$-th step (Fig.\ref{f.4},inset) (reduced
conductance deviation):
\begin{eqnarray}
g_n=\left(2e^{2}/h\right)^{-1}
\frac{\int_{V_{g}(n)}^{V_{g}(n+1)}dV_{g}|\Delta
G(V_{g})|}{V_{g}(n+1)-V_{g}(n)}.
\end{eqnarray}
Inserting here (\ref{approx2}), we get:
\begin{eqnarray}
g_n=\frac{\pi}{2}\sum_{q}\int_{V_{g}(n)}^{V_{g}(n+1)} d V_{g} \aleph_{q}(E_{F})
 \frac{\Gamma_{q}(E_{F},E_{F})}{V_{g}(n+1)-V_{g}(n)},
\end{eqnarray}
where the $n$-th mode opens at the gate voltage equal to $V_{g}(n)$.
In the last equation we can instead of integrating over the gate voltage
integrate over the  subband bottom energies, $E_{\perp}(x)$, which are
determined by $V_{g}$. The intermode distance,  $\Delta E_{\perp}(x)$,
weakly  depends on the gate voltage, the main effect of which was directly
shown to be an  upward shift of the potential   in the constriction
\cite{Cambridge}. In the realistic model of a parabolic confining potential it
also does not depend on the mode number, and in the adiabatic limit is almost
coordinate independent.  Its magnitude is of the order of $E_{F}/N_{max}$,
where $N_{max}$ is the number of transverse  modes in the wide part of the
constriction (and equals approximately to the number of conductance
$2e^2/h$-steps which can be observed in given QPC). This enables us to simplify
the above equation:
\begin{eqnarray}
g_n\approx \frac{\pi}{2}\sum_{q}\int_{E_{q\perp}}^{E_{q\perp}+\Delta
E_{\perp}}\!\!\!\!dE_{\perp}\aleph_{q}(E_{F})
\frac{\Gamma_{q}(E_{F},E_{F})}{\Delta E_{\perp}}. \label{approx3}
\end{eqnarray}
The functions under the integral are implicit functions of the bottom position
of the corresponding mode.

If   localized states    have an almost continuous spectrum (which is
consistent with our initial assumption of the impurity potential being smooth),
 the main dependence on the bottom energy   of the  mode enters the DOS, not
the level width, which is a smooth function of energy in the scale of $\Delta
E_{\perp}$. Therefore we can take it from under the integration. Then we obtain
the
following formula:

\begin{eqnarray}
g_n\approx \frac{\pi}{2}\cdot\frac{\Gamma(E_{F},E_{F};n)}{\Delta E_{\perp}}
 \int_{E_{q\perp}}^{E_{q\perp}+\Delta E_{\perp}}d\!E_{\perp}
\sum_{q}\aleph_{q}(E_{F}). \label{approx4}
\end{eqnarray}
Here we explicitly show that, as follows from the definition (\ref{Gamma}),
$\Gamma_{q}(E_{F},E_{F})$ depends on the number of propagating states in the
contact, that is, on the step number $n$.

The  integral over energy in (\ref{approx4}) has a sense of {\em number of
localized states} in all 1D subbands in the contact, which pass the Fermi level
as we sweep $V_{g}$ across the $n$-th conductance step, $N_Q$.
We will denote its average over possible realizations of the random potential
by $\kappa L$($L$ is the length of the contact):
\begin{eqnarray}
 \kappa = \frac{\langle N_Q \rangle}{L} =
\frac{1}{L}\int_{E_{q\perp}}^{E_{q\perp}+\Delta E_{\perp}} dE_{\perp}  \left<
\sum_{q}\aleph_{q}(E_{F}) \right>,\label{kappa0}
\end{eqnarray}
or
\begin{eqnarray}
\kappa =  \int_{E_{q\perp}}^{E_{q\perp}+\Delta E_{\perp}} dE_{\perp}  \left<
\sum_{q}{\cal N}_q(E_F) \right>.\label{kappa}
\end{eqnarray}
Here $\langle{\cal N}_q(E)\rangle$ is the {\em ensemble average} of DOS {\em
per unit length} in the $q$-th 1D subband over  possible realizations of the
random impurity potential,
so that the ensemble average of $g_n$ is
\begin{equation}
\langle g_n \rangle =
\frac{\pi}{2}\cdot\frac{\left<\Gamma(E_{F},E_{F};n)\right>}{\Delta E_{\perp}}
\kappa L.
\end{equation}

Now the expression (\ref{approx4}) acquires a clear physical meaning. The
correction to the contact conductance due to indirect backscattering is a sum
of contributions from independent Q-states (in each 1D subband):
\begin{equation}
g_n = N_{Q} \Delta g_{n}.
\end{equation}
 Each contribution is approximately the same and is proportional to $n$  (see
(\ref{Gamma})) :
\begin{eqnarray}
\Delta g_{n} = \frac{\pi \langle\Gamma(E_{F},E_{F};n)\rangle}{2 \Delta
E_{\perp}} = \gamma n.
\end{eqnarray}

The density of states is the self-averaging quantity \cite{LGP}, so that in the
limit of infinitely long quantum contact we would get :
\begin{eqnarray}
\lim_{L \rightarrow \infty} \frac{\aleph_q(E)}{L} = \langle{\cal
N}_q(E)\rangle,\\
\nonumber\\
\Rightarrow \lim_{L \rightarrow \infty} \frac{N_Q}{L} =  \kappa.
\end{eqnarray}

 In the contact of finite length the number of appropriate Q-states deviates
from its ensemble average, $\langle N_Q \rangle = \kappa L$, and thus the
conductance of the QPC fluctuates from realization to realization.
In order to predict the performance of a single QPC, we need  the distribution
function of these fluctuations, dependent on such parameters of the contact as
its length $L$ and number of open modes $n$.

 On a length scale large compared to the size of the localized state
(correlation radius of the impurity potential, of order 100 nm), we can regard
these states as {\em independently randomly distributed along the channel, with
occurence per unit length $\kappa$.} The probability to find $p$ such states
along the total length $L$ of the contact is then given  by the Poisson formula
\cite{PROB}:
\begin{equation}
\Pi(N) = \left(\kappa L\right)^{N} e^{-\kappa L}\left/ N! \right..
\end{equation}

Since each state gives the same contribution to the conductance, $\Delta
g_{n} = \gamma n$, then the probability {\em density} of the  conductance
deviation on the $n$-th step

   \begin{eqnarray}
P(g_n = N \gamma n) = \frac{\left(\kappa L\right)^{g/\gamma n} e^{-\kappa
L}}{\gamma n\left(g/\gamma n\right)!}. \label{P2}
\end{eqnarray}

The reduced conductance deviation acquires the discrete set of values, $g_n =
\gamma n, 2 \gamma n, 3 \gamma n ...$, depending on the number of relevant
Q-states in the contact.

The shortcoming of the above formula (discrete set of values of $g_n$) follows
from our simplifying assumption that each Q-state has the same width $\langle
\Gamma(E_F,E_F) \rangle$. In reality, this quantity also fluctuates, so that
the variable $g_n$ is rather a continuous one. Nevertheless,  as the numerical
calculations show, this  expression provides a fairly good description of  the
performance of a QPC (see Sec.\ref{s.A}).

\section{Estimates of the theory parameters}\label{s.3}

We see that the statistics of the conductance fluctuations due to indirect
backscattering is determined by the following parameters (see Eq.(\ref{P2})):
effective length of the QPC, $L$, number of open modes, $n$, average 1D density
of Q-states at the Fermi energy per unit length, $\kappa$, and average
contribution to the conductance from each Q-state, $\Delta g$. Since the latter
quantity is proportional to the average scattering rate from these states to
the propagating ones (see (\ref{Gamma})),
\begin{equation}
\Delta g = \frac{\pi}{2}\frac{\langle\Gamma(E_{F},E_{F};n)\rangle}{\Delta
E_{\perp}} = \gamma n,
\label{gamman}
\end{equation}
it is more convenient to use the set of four independent parameters, $L, n,
\kappa$ and $\gamma$.

The first two are {\em tunable parameters}, which can be changed by changing
$V_{g}$ and/or gate configuration, while the other two are essentially
determined by the properties of given GaAs structure (density and charge of
impurities, spacer thickness etc.), or, in short, by the  impurity potential in
the system.  We need some numerical estimates for   these {\em intrinsic
parameters}, $\gamma$ and $\kappa$. They can be obtained from such
characteristics of the impurity potential as its correlation length, $l_{V}$,
and dispersion, $\sigma^{2} \equiv \langle V_{imp}^{2} \rangle$, which are both
contained in the correlation  function $K({\bf x})$ or  spatial spectral
density $S({\bf k})$ (see \cite{Ziman}):
\begin{eqnarray}
K({\bf x}) = \langle V_{imp}(0)V_{imp}({\bf x})\rangle; \:\:\:\:
S({\bf k}) = \int d^{d}{\bf x} K({\bf x}) e^{-i{\bf k \cdot x}}.
\end{eqnarray}
 (We put to zero the average value of the impurity potential.)

First we estimate $\gamma$. Since  $\langle\Gamma(E_{F},E_{F})\rangle \simeq
2\pi \cdot n \cdot 2\nu \overline{|(Q|V|P)|^{2}}$ (see (\ref{Gamma})), then
\begin{eqnarray}
\gamma \simeq 2\pi^2 \nu \overline{|(Q|V|P)|^{2}}/\Delta E_{\perp}.
\end{eqnarray}
Here $\overline{|(Q|V|P)|^{2}}$ is the average square modulus of the impurity
potential between quasilocalized and propagating states. Evidently, it is of
the order of $S(\Delta k_{PQ})\alpha^{2}/l_{0}$, where $\Delta k_{PQ}$ is the
difference of   longitudinal wave vectors in P- and Q- states,  $\alpha$ is the
reduced matrix element of impurity potential between different transverse modes
($\alpha \simeq 0.1$ according to our numerical calculations, see
Sec.\ref{s.4}), and $l_{0} \simeq l_{V}$ is the average longitudinal size of
the quasilocalized state (the latter factor appears because the wave functions
of Q-states are normalized to unit {\em probability}, while the  P-states are
normalized  to unit {\em probability current} along the axis of the QPC). In
accordance with our basic  assumptions, for the relevant (Q,P)-matrix elements
of the impurity potential the longitudinal wavevectors are almost the same:
$\Delta k_{PQ} \ll k_{F}.$  Therefore we can take
\begin{eqnarray}
\overline{|(Q|V|P)|^{2}} \simeq  S(0)e^{-2\Delta k_{PQ}l_{V}}\alpha^{2}/l_{V} =
S(0)e^{-4\pi\frac{l_{V}}{\lambda_{F}}\frac{\Delta k_{PQ}}{k_{F}}}\alpha^{2}/l_V
 \simeq \sigma^{2} l_{V} \alpha^{2}. \label{dkpq}
\end{eqnarray}
This gives the following estimate for $\gamma$:
\begin{eqnarray}
\gamma   \leq 2\pi^2\alpha^{2} \frac{ \sigma^{2} }{E_{F}\Delta
E_{\perp}}\cdot\frac{l_{V}}{\lambda_{F}}. \label{gamma1}
\end{eqnarray}
We use here the formula for 1D DOS of propagating states, $\nu = 1/\pi\hbar
v_{F}.$

If we insert into (\ref{gamma1}) the values consistent with our numerical
calculations, $\alpha = 0.1$, $\lambda_{F} =  40$nm, $l_{V} = 100$nm, $
\sigma^{2} = 0.1 E_{F}^{2}$, $\Delta E_{\perp} = 0.3 E_{F}$, then we obtain
$\gamma \leq 0.15$. This is a proper order of magnitude estimate, though
higher than the value that follows from the numerical calculations (see
Sec.\ref{s.A}). The reason is, that putting $\Delta k_{PQ} = 0$ in
(\ref{dkpq}) we overestimated $\gamma$. To eliminate this discrepancy, it is
sufficient to take $\Delta k_{PQ}/k_{F} \simeq 7\%$, which is quite reasonable.

One useful remark: In the superposition approximation (see, e.g.,
\cite{Ziman})
we can present the impurity potential as a sum of potentials of identical
potential centres, $v(x)$, with density $n_{imp}$ randomly distributed along
the conducting plane. Then by the   Carson theorem we get  \cite{PROB} $S(k) =
n_{imp}|v(k)|^{2}.$   Therefore the parameter $\gamma$ is proportional to the
impurity density.

  In order to  estimate the other intrinsic parameter, $\kappa$, we use the
quasiclassical, or Thomas-Fermi, approximation for the  average density of
localized states in the 1D subband, $\langle{\cal N}(E)\rangle$ (see
\cite{Ziman}), leading to the following approximate expression:
\begin{eqnarray}
\langle{\cal N}_{q}(E,x)\rangle =
\frac{\sqrt{m}}{\pi\hbar\sqrt{\sigma}}\theta(E_{q\perp}(x)-E)
\exp\left(-\frac{\left(E_{q\perp}(x)-E\right)^{2}}{4\sigma^{2}}\right) {\rm
D}_{-1/2}\left(\frac{ E_{q\perp}(x)-E}{\sigma}\right) \approx \nonumber\\
\approx \frac{\theta(E_{q\perp}(x)-E)}{\sqrt{E_{q\perp}(x)-E}}
\frac{\sqrt{m}}{\pi\hbar}
\exp\left(-\frac{\left(E_{q\perp}(x)-E\right)^{2}}{2\sigma^{2}}\right)
.\label{DOSA}
\end{eqnarray}
D$_{-1/2}(z)$ is the parabolic cylinder function. As a matter of fact, the
validity limits of the latter expression  are determined not by  the
quasiclassicity  conditions, but by much looser ones \cite{LGP}:
\begin{equation}
 |E-E_{q\perp}(x)| \gg \sigma, E_F\cdot \frac{\lambda_F^2}{\l_V^2}.
\end{equation}
Substituting it into the definition of $\kappa$ (\ref{kappa}), we find:
\begin{eqnarray}
\kappa \approx \sum_{q} \int_{E_{q\perp}}^{E_{q\perp}+\Delta E_{\perp}} dE
\frac{\theta(E-E_{F})}{\sqrt{E-E_{F}}} \frac{\sqrt{m}}{\pi\hbar}
e^{-\frac{\left(E-E_{F}\right)^{2}}{2\sigma^{2}}} \nonumber\\
\approx \int_{E_{F}+\Delta E_{\perp}}^{\infty} \frac{dE}{\Delta E_{\perp}}
\Delta E_{\perp}  \frac{1}{\sqrt{E-E_{F}}} \frac{\sqrt{m}}{\pi\hbar}
e^{-\frac{\left(E-E_{F}\right)^{2}}{2\sigma^{2}}}.
\end{eqnarray}
The latter formula can be identically rewritten as follows:
\begin{eqnarray}
\kappa \approx  2^{-1/4} {\bf \Gamma}\left(\frac{1}{4},
\frac{1}{2}\left(\frac{\Delta E_{\perp}}{\sigma}\right)^2\right)\cdot \sqrt{
\frac{\sigma}{E_{F}} } \cdot \frac{1}{\lambda_{F}}. \label{43}
\end{eqnarray}
Here ${\bf \Gamma}(\alpha,z) = \int_{z}^\infty dy\: y^{1-\alpha} e^{-y}$ is the
incomplete gamma function.


With the same choice of parameter values as above, we find that $\kappa \simeq
\sqrt{0.66}$ $\bf{\Gamma}$$(1/4, 1/2)/\lambda_{F} = 0.269/\lambda_{F} \approx
0.006\:\:{\rm nm}^{-1}$.


\section{Numerical calculations}\label{s.4}

 The basis for our hypothesis about the statistics of conductance deviations
 from ideal quantization  is the numerical calculations of QPC conductance
for different realizations of soft random impurity potential, at different
contact lengths and for different number of open modes.

The model used for simulation of a heterostructure device
is very similar to the one
considered in \cite{Davies}.
Donors are assumed to be fully ionized and distributed randomly through the
donor layer.
We restricted donors to   a plane which should be considered as the
middle of the donor layer.
In our calculations all the donors at the $n$- type Al$_{x}$Ga$_{1-x}$As-
doped layer have the same height $h$=30 nm
above the 2DEG.
The electrons are treated as a two-dimensional layer of 10 nm thickness,
which is much smaller than other relevant length scales.
The positions of the impurities were generated by the uniform random
number generator
at the square  1290$\times$1290 nm$^2$.
We assumed periodic boundary conditions, i.e. this square was continued
periodically in all directions in 2D plane.
In order to avoid the occasional appearance of too dense clusters of
impurities at the donor plane the distribution of impurities has been
'relaxed', namely,
we did not allow the appearance of more than 5 impurity atoms
at the area $1/n$, where $n$ is the 2D concentration of impurities.
For the realistic concentration of impurities, $n$ = 10$^{12}$ cm$^{-2}$
one must discard not more than 5 $\div$ 10 $\%$ of impurities.
The potential of a single impurity was taken in the following form:

\begin{equation}
v({\bf r}-{\bf r}_i)= \frac{e}{\varepsilon \sqrt{( {\bf r}-{\bf r}_i)^2+h^2}},
\end{equation}
where $\varepsilon$ ($\approx$ 13) is the static dielectric function of GaAs ,
${\bf r}$ is the position vector
in the 2D plane, and ${\bf r}_i$ is the coordinate of i-th impurity.
Since we did not want to be too specific in modelling any definite
heterostructure,
 we did not take into account the image term
considered in \cite{Davies}.

The potentials of the impurities were summed directly.
The numerical calculations show that the summation of the potentials
of impurities situated within the radius $|{\bf r}-{\bf r}_i| \leq r_{max}=10
h$
gives the resulting potential with a good
precision.
For the rest of the plane the summation could be substituted by integration
over a plane with  homogeneously distributed charge. This gives a constant
term, which can be dropped since we are only interested in the fluctuations of
the impurity potential around its average value.
 Therefore, we choose the average value of the summed potential
as a point of reference for the energy, i.e. we take $\langle V_{imp}\rangle$ =
0.
Further increase of $r_{max}$ does not give any significant change
for the potential fluctuations defined in such a way.

The amplitude of fluctuations
of the {\em unscreened} impurity potential
proves to be  too large,
for a realistic concentration of impurities $n \sim$ 10$^{12}$ cm$^{-2}$
being a few times
greater than the Fermi energy (of the order of $10^{-2}$ eV).
This means that
the screening of this potential by the electrons in 2DEG
should  necessarily be taken into account.
For a qualitative estimation of this effect
we used the way proposed in \cite{Davies}: the Thomas-Fermi approximation,
which is applicable in the case of slowly varying (on the scale of $\lambda_F$)
impurity potential, i.e.,  our case. The density of
electrons is then given by the local equation

\begin{equation}
n({\bf r})=\frac{m^{*}}{\pi \hbar^2} (\mu-eV_{tot}({\bf r}))
\theta(\mu-eV_{tot}({\bf r})), \label{Poisson1}
\end{equation}
where $\mu$ is the chemical
potential, and
$V_{tot}$ is a sum of the unscreened impurity potential and the induced
potential
from the electrons in 2DEG,

\begin{equation}
V_{tot}= V_{imp} + V_{ind}. \label{Poisson2}
\end{equation}
$V_{ind}$ is related to  $n_{ind}$  by the Poisson equation.
For a fixed chemical potential Eqs. (\ref{Poisson1}) and (\ref{Poisson2}) give
a possibility to find
the total potential $V_{tot}$ in a self-consistent manner.
One must keep in mind that
the experimentally observed Fermi energy should be
measured from  the average total potential, namely,
one more equation should be added to this self-consistent scheme,

\begin{equation}
E_F=\mu - \langle V_{tot}\rangle. \label{Poisson3}
\end{equation}
Here $\langle V_{tot}\rangle$ is the average total potential,
$E_F$ is considered as a fixed parameter given by the experiment,
$\mu$ is the "bare" chemical potential.
It shows the degree of filling of the system by electrons
and  varies from iteration to iteration.
Obviously, this system of equations has a single self-consistent solution.
The numerical calculations show that
for $n \sim 10^{12}$ cm$^{-2}$ and $E_F \simeq 10^{-2}$ eV
this solution is achieved when the system is
almost completely filled by the electrons, i.e.,
for the most of the points in 2D plane $n_{ind}$ is non-zero and
the resulting potential fluctuations are smaller than
(or of the order of) the Fermi energy,
$V_{tot} - \langle V_{tot} \rangle \leq E_F$ (see Fig.\ref{f.1}.)
The characteristic scale of the change of the potential fluctuation
is of the order of 100$\div$200 nm.

The
calculations of the potential were made with a  crude mesh in 2D plane.
To solve the scattering problem by the standard
transfer matrix method, a finer mesh is desirable,
and the potential was therefore interpolated by the cubic splines when
necessary.

In all these considerations
we did not take into account the self-consistent effects of redistribution of
electrons
in the constriction due to the gate voltage.
Nevertheless, it is clear that such  effects could be very significant,
since  it could enhance the potential of a single impurity located in a bottle
neck
  because the concentration of electrons in this area is
lower than the average one, and screening is suppressed (impurity denudation
effect).
Such an effect  increases the potential fluctuations at the bottle
neck and can  change significantly the whole behaviour of
the conductance (locking the channel)\cite{Davies}.
The question of how this effect of just a few impurities
located at the narrowest place may show up in the statistics of conductance
fluctuations   is still open.
We plan to discuss such a challenging subject  elsewhere.

In order to obtain the transmittance and reflectance matrices
we solved the 2D Schr\"{o}dinger equation
rewriting it in  matrix form
by the standard transfer matrix method with the proper boundary conditions.
In order to  model  the gate voltage we  used a parabolic
potential, as in \cite{Chao},

\begin{equation}
U_g(x,y)=\hbar^2\eta^2/2m^{*}[2y/b(x)]^2,
\end{equation}
where $\eta=1/4 k_F b(x)$, and

\begin{equation}
b(x)=b_{\infty}-(b_{\infty}-b_0) \sin^2(\pi x/L);\:\:\:\:\:\:-L/2 \leq x \leq
L/2.
\end{equation}
We chose the parameters $b_0$=10 nm and $b_{\infty}$=170 nm, i.e.,
the maximum number of the modes passing through the constriction is eight,
for the realistic $\lambda_F=42$ nm.



As long as we are interested mostly in long constrictions (of length
600 nm and longer) this impurity scattering term will be dominating in
comparison with effects of changes of the width $b(x)$ along the constriction,
a case
thoroughly considered by Brataas and Chao \cite{Chao} for short contacts.
In the bulk of our calculations
we neglect them.
Because the most important region
for the breakdown of quantization lies near the bottleneck,
we multiply the impurity potential inside the contact by the factor $\sin^2(\pi
x/L)$.
This gives us the freedom not to care about changing
the boundary conditions for any realization of the potential.
The number of entering modes is always constant.

The statistics of conductance deviations in quantum contacts
can be obtained  either (i) by generating the impurity potential
(with the self-consistent procedure) for each realization of the constriction,
or (ii) by moving the center of the contact  (possibly also changing   the
direction of
the axis of the contact) over the plane with the impurity potential obtained
in a self-consistent manner only once.
The second way is surely less time consuming and, therefore,
 is the one we use.

We performed numerical calculations
for  contacts of three lengths (600, 800 and 1200 nm) for 625 different
realizations of impurity potential, moving the center of the X-oriented
contact along the square of   1290$\times$1290 nm$^2$, with the periodic
boundary conditions described above.

We also calculated the conductance of very long QPCs (up to 10000 nm)
for several realizations.

\section{Analysis of the results}\label{s.A}

The results of the calculations are shown in Figs.\ref{f.4},\ref{f.4a} and
\ref{f.5}.
Typical   conductance vs. gate voltage curves are shown in Fig.\ref{f.4}.
Curves A correspond to an impurity free contact. The center of the contact lies
at the same place for all the curves B (at a minimum (valley) of the impurity
potential) and C (at a maximum (hill)).    Evidently the C-curves  demonstrate
better quantization than the  B-curves, for the same contact length and step
number. This agrees with our statement that the indirect backscattering is more
effective mechanism of the quantization breakdown in QPCs. The quantization
quality becomes poorer for larger contact length and higher step number.

 The conductance vs. gate voltage curves in a very long QPC (L=10000 nm) for
three different realizations are shown in Fig.\ref{f.4a}.
We would like to note here that no trace of localization in the system is
 detected. On the other hand, the
 conductance plateaus (except the first one) are totally destroyed at this
 length.

In order to obtain more convincing proof of our theory, the statistics of
conductance fluctuations was investigated. The average conductance deviation
from the ideal   conductance step , $g$, was calculated in 625 realizations for
each step ($n=2, 3, .., 6$) and each contact length ($L=600, 800, 1200$nm), and
the empirical distribution function was obtained:
\begin{equation}
P_{emp}(g) \delta g =  N(p\delta g)/625,\label{def}
\end{equation}
where $N(p \delta g)$ is the number of realizations for which $p\delta g \leq g
< (p+1)\delta g$.
In the actual calculations $\delta g = 0.02$.

Since the contact has a finite length,  the deviation $g$ is finite even
without impurities due to the geometric step smoothening \cite{ZK}; it is the
larger the shorter the contact is. Therefore when calculating $g$, we have
substracted from it the corresponding value of $g$ for the   impurity free
case.

The average deviation increases with the mode number and the contact length.
The empirical distributions  are essentially asymmetric. Some of them have
a tail stretching into the negative $g$ region. This means that in some
cases the impurity potential can make the contact more transparent
in comparison with the case without impurities. There may be two mechanisms for
this transparency growth. First, there may occur tunneling through the
localized states in the transition regime, which enhances the contact
transparency \cite{Lev}; second, the random impurity potential may effectively
lessen the aperture of the contact, thus making the conductance steps sharper
and enlarging $g$ \cite{ZK}. Both mechanisms are consistent with the fact that
the negative tails disappear for larger contact length and/or step number, and
both processes are likely to occur in different random potential arrangements.

The values of $g$ obtained from the numerical calculations are not quantized,
as the expression  (\ref{P2}) implies. This is, as stated above, due to the
fact that the widths of different quasilocalized states are not exactly the
same. Nevertheless, by interpolating this formula to noninteger values of $g$
we
achieve a strikingly good description of the numerical results:
\begin{equation}
P_n(g) = \frac{(\kappa L)^{g/\gamma n} e^{-\kappa L}}{\gamma n {\bf
\Gamma}((g/\gamma n) + 1)}. \label{P22}
\end{equation}
This result can be easily understood, if we take into account that the
contribution of each single Q-state to the  correction to the conductance is
small enough ( a few percent of the  total effect, see below), so that a
comparatively large number of the indirect backscattering processes is
necessary to obtain the average effect in the QPC. Then the discrete
distribution (\ref{P2})  almost coincides with its continuous interpolation
(\ref{P22}). This is the more true, the longer the contact is. Indeed, the
fitting of the two curves is better for the larger values of $L$ (see
Fig.\ref{f.5}).

The solid lines in Fig.\ref{f.5} are the least square fits of Eq.(\ref{P22}) to
the empirical  distribution. The dots on these lines indicate the points where
$g$ is an integer multiple of   $\gamma n$ (according to (\ref{P2}). We use the
same value of the parameter $\kappa=(1/320)$ nm$^{-1}$ = $0.13/\lambda_F$ for
all the curves, while the parameter $\gamma$ varied (the mean square error of
the fitting was in the most cases  practically the same as if we would vary
them both independently). The choice of $\kappa$ agrees very well with the
estimates of Sec.\ref{s.3} ( $\kappa \approx 0.269/\lambda_F \approx
0.006\:{\rm nm}^{-1}$) , up to a factor of 2. This shows that the effective
contact length where the scattering occurs is of the order of $L/2$, i.e.,
coincides with the length of its bottleneck part (see Fig.\ref{f.1}).

The values of $\gamma$ are shown in Table \ref{t.1}. They change
insignificantly with $L$ and $n$, and are in a good agreement with the
estimates of Sec.\ref{s.3}. The theoretical curves provide very good
description of both position and magnitude of the distribution peak, as well as
of its large deviation tail.

It is noteworthy  that the width of the resonant peaks in $G(V_{g})$ curves
(Fig.\ref{f.4}) agrees with our estimates of the resonance width $\Gamma \simeq
\gamma n \Delta E_{\perp}$ ($\Delta E_{\perp}$ being of the order of the
stepwidth).

The knowledge of the distribution function of conductance fluctuations allows
to obtain more accurate criteria of good performance of a QPC, e.g., to predict
the probability for a QPC  of a given size, realized in the GaAs structure with
given properties, to have a certain number of well-defined conductance steps.

The analysis of the theory parameters $\gamma$ and $\kappa$ shows that for
different experimental situations   they can be estimated as follows:
\begin{eqnarray}
\gamma \simeq 0.016\cdot\frac{n_{imp}}{n_0};\\
\nonumber\\
\kappa ({\rm nm}^{-1}) \simeq \frac{1}{320}\cdot
\exp\left(1 - \frac{n_0}{n_{imp}}\left(\frac{\Delta E_{\perp}}{\Delta
E_{\perp,0}}\right)^2\right).
\end{eqnarray}
Here $n_{imp},\Delta E_{\perp}$ are the actual parameters of the system, and
$n_0 = 10^{12}$ cm$^{-2}$, $\Delta E_{\perp,0} = 4.1$ meV \cite{FNX} are the
model parameters in our calculations.

The simplest criterion of the QPC performance is given by average deviation
  (see (\ref{P2}))
\begin{equation}
\langle g_n \rangle = \kappa \gamma n  L \simeq 5\cdot 10^{-5} n L (\:{\rm
nm}). \label{13}
\end{equation}
For the channel length $L = 600 (800, 1200)$ nm and for a criterion of total
quantization destruction $\langle g \rangle = 0.5$, this gives $n_{max} \approx
17 (13, 8),$ respectively. The analysis of conductance curves shows that the
quantization is really destroyed at lower values of $\langle g \rangle \simeq
0.2$, i.e. $n_{max} \approx 7 (5, 3)$, which is close to what we see in
Fig.\ref{f.4}.

The transport and localization lengths can be estimated from (\ref{13}) as
follows: $$l_{tr} \approx   (\kappa \gamma n)^{-1}; \:\: l_{loc} \geq
l_{tr}\cdot n.$$  This leads to the following  results: for $n$ open modes
$ l_{tr} \approx 20 000/n$(nm);  $ l_{loc} \geq 20 000$(nm). These values are
much larger than experimentally reasonable contact lengths; they justify our
theoretical approach of Sec.\ref{s.1}.

In the recent experiment \cite{Koester} the quantized conductance was observed
in an InAs/AlSb ballistic constriction with channel length 1000 nm. Up to eight
conductance steps were detected. Koester et al. suggest that much better
performance of their device  compared to  ones utilizing
GaAs/Al$_x$Ga$_{1-x}$As
heterostructures  is due to higher ratio of  interlevel energy spacing to the
amplitude of the impurity potential fluctuations. This is in a complete
agreement with our theory, since the density of quasilocalized states
(and the parameter $\kappa$ of the theory, see Eq.(\ref{43})) exponentially
drops with growing $\Delta E_{\perp}/\sigma$.

\section{Conclusion}\label{s.5}

In conclusion, we have investigated the effects of electronic scattering by the
soft
impurity potential in quantum point contacts both with use of an analytical
model and by direct numerical calculations.

We have shown that the  decisive mechanism of conductance quantization
breakdown is due to the
indirect backscattering of carriers via quasilocalized states at the Fermi
level. For the realistic contact lengths they can be described in terms of
independent scatterers, though the electron propagation is coherent. This
is due to the smoothness of the scattering potentials, leading to very large
scattering and localization lengths.

The performance of the quantum contact is shown to be strictly dependent on the
ratio of  the intermode  distance to the
amplitude of the random impurity potential (the larger the better).

For the first time we have obtained  analytical and empirical formulas
both for the conductance deviation
due to  this process and for the probability distribution of these deviations
in an ensemble of contacts with realistic potentials.
The latter has proved to be a generalized Poisson distribution.

The  parameters
of the distribution obtained numerically agree quite
well with  analytical calculations based on general assumptions, thus
confirming their
applicability to the case of quantum transport through the QPC in the presence
of the random impurity potential.

\acknowledgements
We are grateful to Mats Jonson and Ilya Krive for fruitful discussions.

\begin{table}
\begin{tabular}{||l||c|c|c|c|c||l||}
L(nm)$\backslash$n & 2 & 3 & 4 & 5 & 6 & $\bar{\gamma}(L)$ \\
\hline
\hline
600 & 0.019 & 0.018 & 0.018 & 0.017 & 0.016 & 0.018 \\
\hline
800 & 0.016 & 0.017 & 0.017 & 0.016 & 0.015 & 0.016 \\
\hline
1200 & 0.014 & 0.014 & 0.014 & 0.013 & 0.012 & 0.014 \\
\hline
\hline
$\bar{\gamma}(n)$& 0.016 & 0.016 & 0.016 & 0.015 & 0.014 & $<\gamma>=$ 0.016\\
\end{tabular}
\caption{ Parameter $\gamma$ as determined by the least square fitting of
Eq.(51) to the numerical data for different contact lengths $L$ and step
numbers $n$. Parameter $\kappa$ is kept equal to (1/320) nm$^{-1}$ for all
$L,n$.}\label{t.1}
\end{table}

\begin{figure}
\caption{The model of a quantum point contact.  The impurity potential in  2DEG
layer. Impurity density is 10$^{12}$ cm$^{-2}$. Isolines are drawn through
$1.3\times 10^{-3}$ eV. The  white line shows the gate equipotential
$U_{g}^{(0)}(x,y)=E_{F}$ (unperturbed contact shape) for $L=800$nm.   }
\label{f.1}
\end{figure}

\begin{figure}
\caption{Classification of electronic states in a QPC. $E_{m\perp}(x)$ -
effective potential in $m$-th mode (1D subband). $(\alpha = -2)$ - reflected
states incident from the left;  $(\alpha = -1)$ - propagating states incident
from the left; $(\alpha = 0)$ - quasilocalized states; $(\alpha = 1)$ -
propagating states incident from the right; $(\alpha = 2)$ - reflected states
incident from the right.}
\label{f.2}
\end{figure}


\begin{figure}
\caption{Conductance vs. gate voltage curves for different realizations and
contact lengths (see text). Parameter $Z = k_F b(0)/\pi$. a) - contact length
1200 nm, b) - 800 nm, c) - 600
nm. Inset: Quantization breakdown parameter $g$; it equals to the ratio of the
area of dashed part of the conductance step to its total area. }
\label{f.4}
\end{figure}

\begin{figure}
\caption{Conductance vs. gate voltage curves for three different realizations
for the
contact length L=10 000 nm (see text). The random potential has been generated
on the  rectangle 1290$\times$12900 nm$^2$ for these calculations.}
\label{f.4a}
\end{figure}

\begin{figure}
\caption{Statistics of conductance fluctuations in  QPCs (see text). a) -
contact length 1200 nm, b) - 800 nm, c) - 600 nm.  }
\label{f.5}
\end{figure}


\begin{thebibliography}{99}
\bibitem{FNX} In fact, the value of $\Delta E_\perp = 4.1$ meV coincides with
directly measured in \cite{Cambridge} distance between the first and second
transverse modes, which ranged from 4.0 to 4.6 meV for different gate voltages.
\end{thebibliography}
\end{document}